\documentclass[aps,prb,preprint,superscriptaddress]{revtex4-2}

\usepackage{graphicx}
\usepackage{color}

\begin{document}



\title{Simultaneous collapse of antiferroquadrupolar order and superconductivity in PrIr$_{2}$Zn$_{20}$ by nonhydrostatic pressure}



\author{Kazunori Umeo}
\email[e-mail address: ]{kumeo@hiroshima-u.ac.jp}
\affiliation{Department of Low Temperature Experiment, Integrated Experimental Support / Research Division, N-BARD, Hiroshima University, Higashi-Hiroshima, 739-8526, Japan}
\author{Riho Takikawa}
\affiliation{Graduate School of Advanced Science and Engineering, Hiroshima University, Higashi-Hiroshima 739-8530, Japan}
\author{Takahiro Onimaru}
\affiliation{Graduate School of Advanced Science and Engineering, Hiroshima University, Higashi-Hiroshima 739-8530, Japan}
\author{Makoto Adachi}
\affiliation{Department of Quantum Matter, AdSM, Hiroshima University, Higashi-Hiroshima 739-8530, Japan}
\author{Keisuke T. Matsumoto}
\affiliation{Graduate School of Science and Engineering, Ehime University, Matsuyama, Ehime 790-8577, Japan}
\author{Toshiro Takabatake}
\affiliation{Graduate School of Advanced Science and Engineering, Hiroshima University, Higashi-Hiroshima 739-8530, Japan}


\date{\today}

\begin{abstract}

Superconductivity in PrIr$_{2}$Zn$_{20}$ appears at $T_{\rm c} = 0.05$ K in the 
presence of an antiferroquadrupolar order below $T_{\rm Q} = 0.11$ K. We have 
studied pressure dependences of $T_{\rm c}$, $T_{\rm Q}$, and non-Fermi liquid 
behaviors in the resistivity $\rho (T)$ by using two pressure transmitting 
media: argon maintaining highly hydrostatic pressure, and glycerol, which 
solidifies above 5 GPa producing nonhydrostatic pressure. Upon applying $P$ 
with argon up to 10.6 GPa, $T_{\rm c}$ hardly changes, while $T_{\rm Q}$ monotonically 
increases from 0.11 to 0.23 K. With glycerol, however, $T_{\rm Q}$ and 
$T_{\rm c}$ simultaneously fall below 0.04 K at 6.3 GPa. The contrasting results indicate that onsite quadrupolar fluctuations induce superconductivity in this compound. 

\end{abstract}


\maketitle


\newpage

\section{INTRODUCTION}
Unconventional superconductivity in correlated materials \cite{MR01} such as high-$T_{\rm c}$ copper oxides \cite{TM02}, ruthenium oxides \cite{AP03}, iron-based pnictides  \cite{HH04}, 
and heavy-fermion compounds \cite{CP05} has attracted tremendous attention over 
three decades as one of the most intriguing aspects in condensed-matter physics. 
Fluctuations of magnetic dipole moments in them are thought to be the 
glue for superconducting pairing \cite{MR01,TM02,AP03,HH04,CP05}. In recent years, pairing interaction 
due to quadrupole fluctuations has been proposed in praseodymium-based 
intermetallic compounds \cite{ED06,YA07}. However, the role of the quadrupole freedom 
in superconductivity remains elusive in spite of detailed studies for 
several systems. PrOs$_{4}$Sb$_{12}$, for example, exhibits 
superconductivity below $T_{\rm c}=1.85$ K and a magnetic-field-induced 
antiferroquadrupole (AFQ) order above $B = 4.5$ T \cite{YA07}. The field-induced AFQ 
order results from the crossing of the crystalline electric field (CEF) 
levels between the $\Gamma_{1}$ singlet ground state and the Zeeman-split 
$\Gamma_{4}$ level with quadrupolar freedom \cite{KK08}. Because the 
superconducting phase is located in the vicinity of the quadrupole ordered 
phase, it was argued that the quadrupole fluctuations may play an important 
role in the superconductivity \cite{KK08,KH09}. 

Recently, superconductivity was discovered in the series of compounds 
Pr$T_{2}X_{20}$ ($T =$ Ir, Rh, $X =$ Zn; $T =$ Ti, V, $X =$ Al) \cite{TK10,AS11,TO12}. The local 
point-group symmetry of the Pr site is $T_{\rm d}$ (or $T)$ in the cubic 
CeCr$_{2}$Al$_{20}$-type structure with the space group $Fd\bar{3}m$. The 
CEF ground state of 4$f^{2}$ electrons in the Pr$^{3+}$ ion was identified as a 
non-Kramers $\Gamma_{3}$ (or $\Gamma_{23})$ doublet with quadrupolar 
freedom \cite{TO12}. Interestingly, the superconducting transition sets in below 
the ordering temperature of either antiferroquadrupoles or ferroquadrupoles. In 
Pr$T_{2}$Zn$_{20}$ ($T =$ Ir, Rh), AFQ transitions at $T_{\rm Q}= 0.11$ and 
0.06 K are followed by superconducting ones at $T_{\rm c}= 0.05$ and 
0.06 K, respectively \cite{TO13,TO14}. In PrTi$_{2}$Al$_{20}$, on the other hand, a 
ferroquadrupole (FQ) transition at $T_{\rm Q}= 2$ K is followed by a 
superconducting one at $T_{\rm c}= 0.2$ K \cite{AS15}. The coexistence of 
the AFQ or FQ order with the superconductivity led to the proposition that 
the superconducting pairing interaction is mediated by quadrupole 
fluctuations \cite{TO13,TO14,AS15}. 

In Pr$T_{2}$Zn$_{20}$ ($T =$ Ir, Rh), the electrical resistivity $\rho (T)$ 
decreases with a downward curvature on cooling from 2 K to $T_{\rm Q}$, and the 
$4f$ contribution to the specific heat divided by temperature 
$C_{4f}/T$ follows the form of --ln$T$ for 0.2 $\textless$ $T$ $\textless$ 0.8 K 
\cite{TO16}. These non-Fermi-liquid (NFL) behaviors in $\rho (T)$ and $C_{4f}/T$ are consistent with the predictions from the two-channel Anderson lattice model \cite{AT17}. Furthermore, the values of magnetic entropy at $T_{\rm Q}$ for $T =$ Ir and Rh are, respectively, 20{\%} and 10{\%} of $R$ln2 \cite{TO13,TO14}. The reduced entropy was attributed to fluctuations of the quadrupole moments which persevere temperatures up to
$20 - 30$ times higher than $T_{\rm Q}$. 

Application of pressure can control the CEF level scheme as well as the strength 
of the hybridization between the $4f$ and conduction electrons ($c$-$f$ hybridization) 
which is an important parameter for the quadrupole Kondo effect and the 
Ruderman-Kittel-Kasuya-Yosida (RKKY)-type interaction among quadrupoles. Upon 
applying pressure to PrOs$_{4}$Sb$_{12}$ up to 8 GPa, $T_{\rm c}$ is 
suppressed monotonically, and the field-induced AFQ phase moves to lower 
fields \cite{NK18}. This change in $T_{\rm c}$ was attributed to the decrease in the 
energy splitting between the $\Gamma_{1}$ singlet ground state and 
$\Gamma_{4}$ triplet excited state \cite{NK18}. For PrTi$_{2}$Al$_{20}$, 
$T_{\rm Q}$ begins to be suppressed above 6 GPa, where $T_{\rm c}$ is strongly 
enhanced together with the effective mass of the quasiparticles \cite{KM19}. The 
opposite changes in $T_{\rm Q}$ and $T_{\rm c}$ support the proposition that the enhanced fluctuations of quadrupoles in the vicinity of the quantum critical point strengthen the superconducting interaction.

We recall that the AFQ order is sensitive to the uniformity of the pressure 
applied to the crystal. In fact, an AFQ order at $T_{\rm Q} = 0.4 $K in PrPb$_{3}$ suddenly 
disappears at 5 GPa \cite{MK20}, which coincides with the 
solidification pressure of the transmitting medium of glycerol \cite{AD21}. It is 
likely that the solidification causes nonuniform pressure which lowers the 
local symmetry of Pr$^{3+}$ and lifts the twofold degeneracy of the $\Gamma 
_{3}$ doublet. Therefore, comparing the effects of nonhydrostatic and 
highly hydrostatic pressures on $T_{\rm Q}$ and $T_{\rm c}$ in PrIr$_{2}$Zn$_{20}$ may give a clue for understanding the relation between the 
superconductivity and AFQ fluctuations. Bearing this in mind, we have 
measured the electrical resistivity $\rho (T)$ of PrIr$_{2}$Zn$_{20}$ under 
pressures applied using two types of pressure-transmitting media: glycerol producing 
nonhydrostatic pressure above 5 GPa, and argon, keeping pressure hydrostatic 
up to 10 GPa \cite{SK22,NT23}. We also compare these data with the pressure 
dependence of $T_{\rm c}$ for the BCS superconductor LaIr$_{2}$Zn$_{20}$ \cite{TK10,KT24}. A previous study of the La-substituted system 
Pr$_{1-x}$La$_{x}$Ir$_{2}$Zn$_{20}$ showed that $T_{\rm Q}$ vanishes even at a small $x$ $\textless$ 0.09, while $T_{c}$ hardly changes for a wide range $0 \le x \le 0.47$ \cite{KT24}. The independent behaviors between $T_{\rm Q}$ and $T_{\rm c}$ for the La-substituted system have not disclosed the role of AFQ fluctuations in the 
superconductivity. In this paper, we compare variations of $T_{\rm Q}$ and $T_{\rm c}$ under hydrostatic and nonhydrostatic pressures. The results indicate that on-site coupling between quadrupole fluctuations and conduction electrons induces superconductivity in PrIr$_{2}$Zn$_{20}$. 

\section{EXPERIMENTAL DETAILS}
Single crystals of PrIr$_{2}$Zn$_{20}$ and LaIr$_{2}$Zn$_{20}$ were prepared 
by a melt-growth method from high-purity elements, Pr (4N), La (4N), Ir (3N), 
and Zn (6N), as reported previously \cite{TK10}. The $\rho (T)$ under pressure was 
measured by an ac four-terminal method using a piston-cylinder pressure cell 
for $P \le 2.1$ GPa and an opposed-anvil pressure cell for $P \ge 3.2$ GPa. The pressure was applied at room temperature.
In the piston-cylinder cell, Daphne oils 7373 and 7474 were used as pressure 
transmitting media which produce hydrostatic pressure until they solidify at 2.3 and 3.7 GPa at room temperature, respectively \cite{KM25}. In the opposed-anvil cell, argon and glycerol were used as described above. The pressure was estimated from the 
pressure dependence of $T_{\rm c}$ of a piece of lead placed in the cell \cite{BB26}. The magnitude of pressure gradients $\Delta P$ was estimated from the 
temperature width of the superconducting transition of the Pb manometer. A commercial Cambridge Magnetic Refrigerator mFridge mF-ADR50 was used to cool the pressure cell down to 0.04 K.

\section{RESULTS AND DISCUSSION}
Figures 1(a) and 1(b) represent the low-temperature data of $\rho (T)$ under 
pressures applied using argon and glycerol as pressure-transmitting media, 
respectively. Upon applying pressure with argon, $T_{\rm c}$ in $\rho (T)$ remains 
unchanged, whereas the sharp drop at $T_{\rm Q}$ shifts to higher temperatures. 
The steady increase in $T_{\rm Q}$ suggests the AFQ order is stabilized by 
the increased RKKY-type interaction among the quadrupole moments. On the other hand, upon applying pressure with glycerol, the anomalies at $T_{\rm c}$ and $T_{\rm Q}$ as well as the downward curvature above $T_{\rm Q}$ remain up to 4.9 GPa, where glycerol 
holds a liquid state at room temperature. However, at 6.3 GPa, where glycerol 
solidifies, we did not observe any transition in $\rho (T)$ down to the lowest 
temperature, 0.04 K. It is noted that the downward 
curvature in $\rho (T)$ also disappears as shown in Fig. 1(b). The concomitant
suppression of $T_{\rm c}$ amd $T_{\rm Q}$ and the downward curvature in $\rho 
(T)$ suggest that the scattering of conduction electrons by means of the $\Gamma 
_{3}$ doublet is significantly modified under nonhydrostatic pressure. 
In fact, pressure gradients along the Pb manometer increase slowly from $\Delta P = 0.01$ GPa at 3.2 GPa to 0.02 GPa at 4.9 GPa, but steeply to $\Delta P = 0.39$ GPa at 6.3 GPa as demonstrated in Fig. S4 in the Supplemental Material for the details \cite{SS27} (see, also, references \cite{SS27a, SS27b} therein). This pronounced increase of the pressure gradient in the range 4.9-6.3 GPa should cause a large anisotropic strain in the sample. 

The pressure dependences of $T_{\rm c}$ and $T_{\rm Q}$ are compared in Fig.\ 2 for 
argon and glycerol media together with the data for Daphne oil for $P \le $ 
2.1 GPa (See Fig. S1 in the Supplemental Material for the details \cite{SS27}). $T_{\rm c}$ was defined as the onset temperature of the drop in $\rho (T)$ in Fig. 1, and  $T_{\rm Q}$ was taken as the peak temperature of $d\rho/d$\textit{T}. For $P = 3.2$ and 4.9 GPa, $T_{\rm c}$ and $T_{\rm Q}$ for 
glycerol are slightly lower than those for argon, suggesting the influence of 
the increased viscosity of glycerol near the solidification pressure. On 
further pressurizing with glycerol up to 6.3 GPa, both $T_{\rm c}$ and $T_{\rm Q}$ vanish. The vanishment of $T_{\rm Q}$ is attributed to the possible 
anisotropic strain caused by the nonhydrostatic pressure, which lowers the cubic  symmetry of the Pr site. Thereby, the nonmagnetic $\Gamma_{3}$ 
doublet of the CEF ground state of Pr$^{3+}$ splits and loses the quadrupolar degree of freedom. The concomitant disappearance of $T_{\rm c}$ with $T_{\rm Q}$ is indicative of a strong correlation between the superconductivity and AFQ order. 

Let us compare the above results with those for the substitution of La for Pr in 
Pr$_{1-x}$La$_{x}$Ir$_{2}$Zn$_{20}$ \cite{KT24}. It was found that $T_{\rm Q}$ disappears at a small substitution level $x = 0.09$, while the superconducting transition remains 
in the whole range $0 \le x \le 1$ \cite{KT24}. This fact was interpreted to be an 
indication of the lack of a relation between $T_{\rm c}$ and $T_{\rm Q}$. Now, it is noteworthy 
that the electronic state of Pr$^{3+}$ in the substituted system is 
different from that under pressure. In our previous paper \cite{KT24}, two mechanisms were considered to explain the suppression of $T_{\rm Q}$ in the La-substituted system. First, the expectation value of the quadrupoles is 
reduced when the $\Gamma_{3}$ doublet is split. Second, the intersite RKKY-type quadrupole interaction  \cite{YI29} is weakened by the breaking of coherence in the lattice due to random distribution of Pr and La atoms. 

The second mechanism was considered to play the dominant role in the collapse of 
the AFQ order in the La-substituted system because the first one was discarded. In fact, the magnetic susceptibility and specific heat data for 
Pr$_{1-x}$La$_{x}$Ir$_{2}$Zn$_{20}$ indicated the stability of the $\Gamma 
_{3}$ doublet even in the substituted system  \cite{KT24}. Under this condition, 
quadrupolar fluctuations at each Pr site do not vanish. The quadrupolar fluctuations manifest themselves in the downward curvature of $\rho (T)$ for $x = 0$ on cooling below 2 K due to the quadrupolar Kondo effect \cite{AT17,YY30,TY31}. The 
downward curvature in $\rho (T)$ remains for $x = 0.22$ in which no AFQ 
order occurs, but a superconducting transition appears below 0.07 K \cite{KT32}. Therefore, 
superconductivity survives in the La-substituted system if it originates 
from the on-site coupling between quadrupole fluctuations and conduction 
electrons. 

On the other hand, the first mechanism is responsible for the collapse of 
the AFQ order under non-hydrostatic pressure applied by solid glycerol. Once 
onsite quadrupole fluctuations are quenched, superconductivity no longer 
survives. The disappearance of the downward curvature in $\rho (T)$ at 
6.3 GPa as shown in Fig. 1 (b) strongly indicates the quenching of the 
quadrupole fluctuations. This scenario is consistent with the coexistence of AFQ order and 
superconductivity under hydrostatic pressures applied by argon. Furthermore, 
according to this scenario, superconductivity may survive until the uniaxial stress becomes large enough to stop on-site quadrupole fluctuations. In fact, by using another pressure medium, a Fluorinert 70/77$=1:1$ mixture, we have observed both the AFQ order and superconducting transition up to 9.6 GPa beyond the solidification pressure of 1 GPa as shown in Figs. S2 and S3 in the Supplemental Material\cite{SS27}. The transition happened because the uniaxial stress in the solid of Fluorinert 
was small compared with that of solid glycerol at $P$ $\textgreater$ $6$ GPa (see Sec. 3 of the Supplemental Material for details \cite{SS27}.) All observations for 
PrIr$_{2}$Zn$_{20}$ under hydrostatic and non-hydrostatic pressures as well 
as for the La substituted system are consistent with the idea that on-site 
coupling between quadrupole fluctuations and conduction electrons induces 
superconductivity in PrIr$_{2}$Zn$_{20}$. 

To shed light on the pairing mechanism for the superconductivity, we 
compare the observed pressure effect on PrIr$_{2}$Zn$_{20}$ with theories 
and those on other systems. The above argument is consistent with the 
theoretical calculation showing that the superconductivity for the $f^{2}$ 
state of the $\Gamma_{3}$ non-Kramers doublet system appears only in the 
quadrupole ordered phase \cite{KK34}. In this theory, the on-site pairing state 
composed of the $\Gamma_{3}$ doublet state is indispensable to the 
superconductivity. The pressure dependence of $T_{\rm c}$ of 
PrIr$_{2}$Zn$_{20}$ is distinguished from that of PrOs$_{4}$Sb$_{12}$ 
with the $\Gamma_{1}$ CEF ground state. In the latter, $T_{\rm c}$ of 1.85 K 
at $P = 0$ decreases continuously to 1.3 K by applying pressure up to 8 GPa 
even with the glycerol transmitter. Note that there is no anomaly in 
$T_{\rm c}(P)$ around 5 GPa \cite{NK18}, unlike the case of PrIr$_{2}$Zn$_{20}$. 
Furthermore, as shown in Fig.\ 2(a), under hydrostatic pressure, $T_{\rm c}$ of 
PrIr$_{2}$Zn$_{20}$ hardly changes up to 10 GPa. For PrTi$_{2}$Al$_{20}$, 
$T_{\rm c}$ also hardly changes up to 6 GPa \cite{KM35}. In the BCS-type 
superconductor LaIr$_{2}$Zn$_{20}$, however, $T_{\rm c}$ of 0.6 K at $P = 0$ is 
suppressed with a ratio of $-0.12$ K/GPa, as shown in Fig.\ 3. This 
contrasting behavior in $T_{\rm c}(P)$ between PrIr$_{2}$Zn$_{20}$ and 
LaIr$_{2}$Zn$_{20}$ strongly suggests a non-BCS paring mechanism for the 
superconductivity in PrIr$_{2}$Zn$_{20}$. If onsite interaction between 
quadrupoles and conduction electrons at the Pr site plays an important role 
in the pairing mechanism as proposed by theories \cite{KK34,KK36,KK37}, the 
superconductivity would be robust against hydrostatic pressure. 

We turn our attention to the hydrostatic pressure effect on $\rho (T)$ for $T$ $\textgreater$  
$T_{\rm Q}$. On cooling, $\rho (T)$ at ambient pressure transforms from a 
$T$-linear line to a curve with a downward curvature, as shown in Fig.\ $4(\rm a)$. We 
denote the crossover temperature as $T_{\rm R}$, as indicated by the arrow, which is defined as the temperature where $\rho (T)$ starts deviating from the linear dependence on cooling. A 
theory based on the two-channel Anderson lattice model shows that $T_{\rm R}$ increases as the $c$-$f$ hybridization is increased \cite{AT17}. In fact, an 
experimental study substituting Cd and Ga for Zn in PrIr$_{2}$Zn$_{20}$ 
confirmed that $T_{\rm R}$ is controlled by the strength of $c$-$f$ hybridization \cite{RJ38}. 
Therefore, $T_{\rm R}$ can be regarded as the measure of the strength of the $c$-$f $ hybridization. 
In Fig.\ 4, the data for $\rho (T)$ under various constant pressures show that 
$T_{\rm R}$ increases with $P$ from 1.8 to 2.6 K at around 6 GPa and decreases to 2.1 
K at 10.6 GPa [see Fig.\ 4(b)]. 

Figure 4 (c) displays a plot of $\Delta \rho (T)$/$\Delta \rho 
(T_{\rm R})$ vs $T$/$T_{\rm R}$, where $\Delta \rho (T) = \rho (T) - \rho 
_{\rm 0}$ and $\rho_{\rm 0}$ is the residual resistivity. Note that all the 
data taken under pressures from 0 to 10.6 GPa fall on one curve in a rather 
wide temperature range of 0.1 $\textless$ $T$/$T_{\rm R}$ $\textless$ 1.2. The dashed 
curve is a fit of the following form derived from the two-channel Anderson 
lattice model \cite{AT17}:

\begin{center}
\def\dfrac#1#2{{\displaystyle\frac{#1}{#2}}}
$\Delta \rho =\dfrac{a_{1}}{1+a_{2}{\dfrac{T_{\rm R}}{T}}}$ ,
\end{center}

where $a_{1}$ and $a_{2}$ are parameters. This scaling underlines that the NFL 
behavior in $\rho (T)$ characterized by the downward curvature is the 
manifestation of the quadrupolar Kondo lattice. 

Let us discuss the pressure dependence of $T_{\rm R}$. The increment of $T_{\rm R}$ 
for $P$ $\textless$ 6 GPa is understood as a result of the enhancement of the 
$c$-$f$ hybridization under pressures. However, this relation does not hold for $P$ $\textgreater$ 6 GPa, where $T_{\rm R}$ decreases with pressure. On the other hand, by 
substituting Ga for Zn in PrIr$_{2}$Zn$_{20}$, $T_{\rm R}$ is increased as a 
result of the increase in the 4$p$ electron density of states of the conduction 
bands \cite{RJ38}. Therefore, the decrease in $T_{\rm R}$ for $P$ $\textgreater$ 6 GPa may 
result from the decrease of the density of states at the Fermi level. 
Therefore, the pressure-induced enhancement of $c$-$f$ hybridization gives way to 
suppression of $c$-$f$ hybridization by the decrease of the density of states. To 
confirm this scenario, the modification of the band structure under 
pressure needs to be calculated.

\section{CONCLUSION}
In iron-based and heavy-fermion superconductors, magnetic fluctuations 
and/or orbital fluctuations have been discussed as the Cooper pairing 
interaction thus far, despite the prevailing controversy on the detailed 
mechanism \cite{TM02,HH04,CP05,ED06,YA07,KK08,KH09,SO39,AK40,TN41}. In this work, we have measured the electrical 
resistivity of PrIr$_{2}$Zn$_{20}$ under pressure applied using two different 
pressure-transmitting media, argon (hydrostatic) and glycerol 
(nonhydrostatic for $P \textgreater 5$GPa). Under hydrostatic pressure, $T_{\rm c}$ hardly changes up 
to 10.6 GPa, while $T_{\rm c}$ for the isostructural BCS superconductor 
LaIr$_{2}$Zn$_{20}$ largely decreases. This contrasting pressure dependence 
of $T_{\rm c}$ corroborates the unconventional nature of the superconductivity in 
PrIr$_{2}$Zn$_{20}$. On the other hand, the AFQ order is stabilized as 
manifested by the monotonic increase in $T_{\rm Q}$ from 0.11 to 0.23 K. 
Under nonhydrostatic pressure at $P = 6.3$ GPa, both $T_{\rm Q}$ and $T_{\rm c}$ 
simultaneously disappear. The comparison between the 
pressure effect and substitution effect on $T_{Q}$ and $T_{c}$ strongly 
suggests that on-site coupling between quadrupole fluctuations and conduction 
electrons induces superconductivity in PrIr$_{2}$Zn$_{20}$. Furthermore, NFL behavior in $\rho $($T$ $\textgreater$ $T_{\rm Q})$ with the downward curvature also disappears under nonhydrostatic pressure. These observations reveal that the NFL behavior as 
well as the superconductivity in PrIr$_{2}$Zn$_{20}$ results from the 
quadrupolar degree of freedom through $c$-$f$ hybridization. Our finding will pave 
the way to a deeper understanding of the pairing mechanism mediated by 
orbital fluctuations in iron-based and heavy-fermion superconductors.

\begin{acknowledgments}

We acknowledge valuable discussions with K. Kubo, K. Matsubayashi, K. Izawa, 
A. Tsuruta, K. Miyake, Y. Shimura, Y. Yamane, and R. Yamamoto. We also thank S. Hara, K. Ohfuka, Y. Sugano, and T. Ohsuka for their technical support. The 
resistivity measurements under high pressures at low temperatures were 
performed at N-BARD, Hiroshima University. This work was partly supported by 
Japan Society for the Promotion of Science KAKENHI Grants No. JP25400375, No. JP26707017, No. JP15H05886, No. JP18H01182, and No. JP18K03518.

\end{acknowledgments}

\bibliography{basename of .bib file}

\newpage

%


\begin{figure}
\includegraphics[width=10cm]{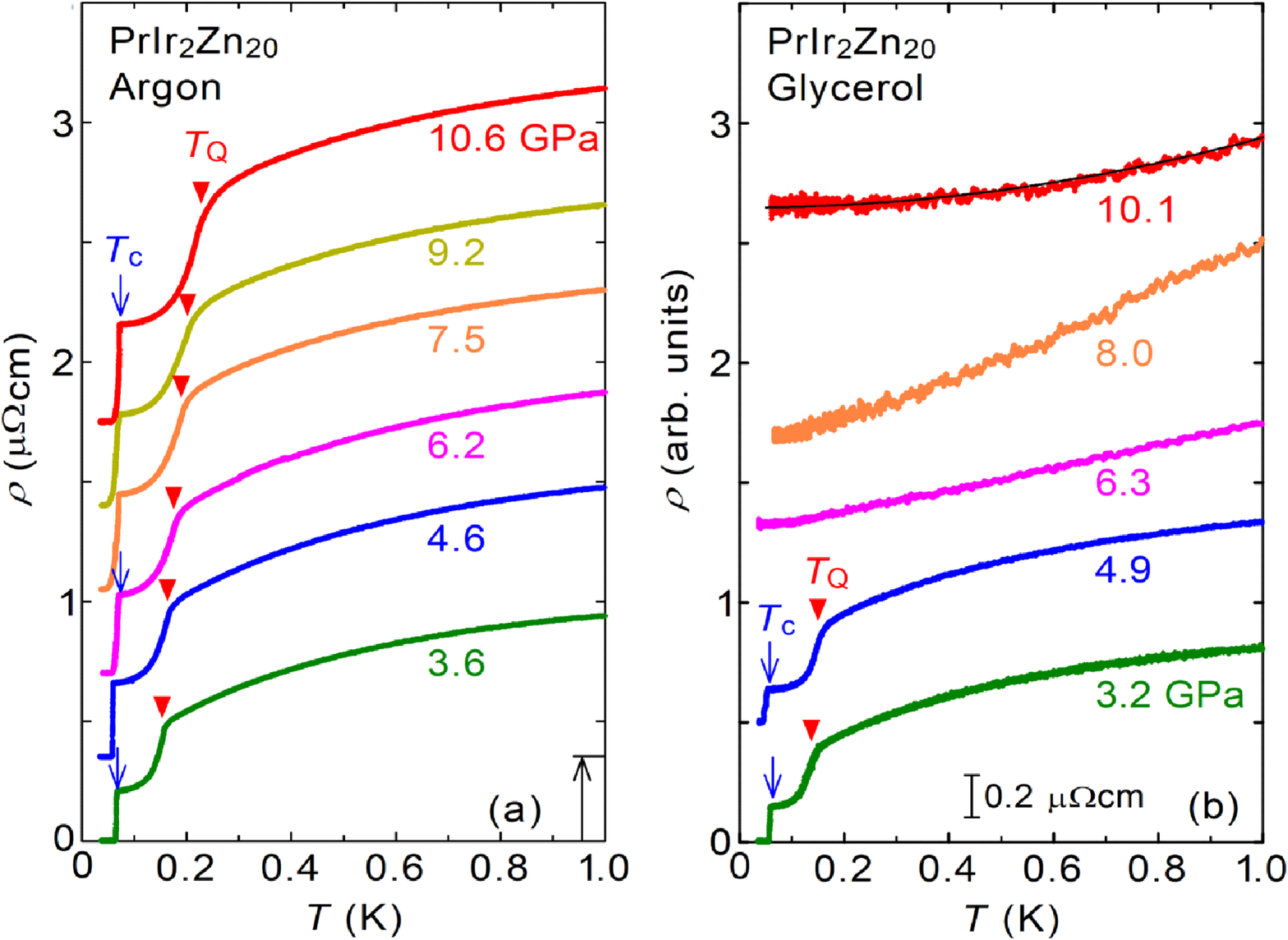}%
\caption{\label{fig:epsart}Temperature dependence of the electrical resistivity 
$\rho (T)$ of PrIr$_{2}$Zn$_{20}$ under various pressures applied using (a) argon and (b) 
glycerol as pressure-transmitting media. The triangles indicate 
the AFQ ordering temperature $T_{\rm Q}$. The data sets at various pressures are 
shifted upward consecutively for clarity. The solid curve for $P = 10.1$ GPa 
represents the fit with $\rho (T) = \rho_{0}+$ \textit{AT}$^{2}$, with $A =$ 
0.3 $\mu \Omega $ cm/K$^{2}$. This Fermi-liquid behavior in $\rho (T)$ suggests that 
the splitting of the $\Gamma_{3}$ doublet stabilizes the Fermi-liquid state.}
\end{figure}

\begin{figure}
\includegraphics[width=10cm]{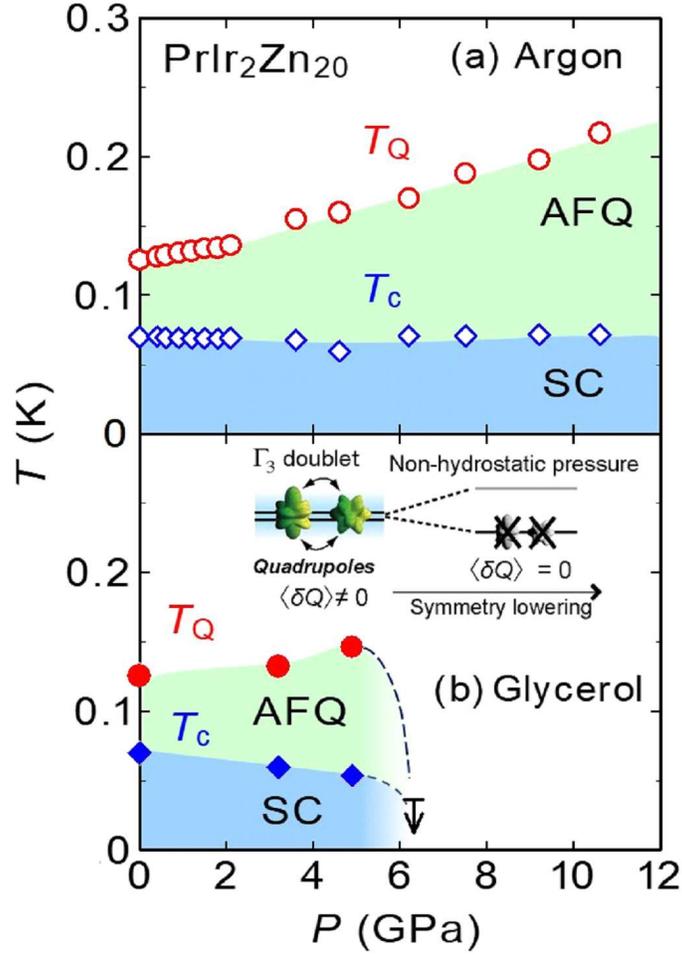}%
\caption{\label{fig:epsart}Pressure dependences of the superconducting transition 
temperature $T_{\rm c}$ and AFQ ordering temperature $T_{\rm Q}$ using (a) argon and (b) 
glycerol as the pressure-transmitting media, respectively. The data 
for $P \le  2.1$ GPa in Fig. 2(a) are taken using a piston-cylinder cell with Daphne oil 7474  (see Fig. S1 in the Supplemental Material for details \cite{SS27}). The inset in (b) represents a schematic diagram for the non-hydrostatic effect on the $\Gamma_{3}$ doublet. 
}
\end{figure}

\newpage

\begin{figure}
\includegraphics[width=10cm]{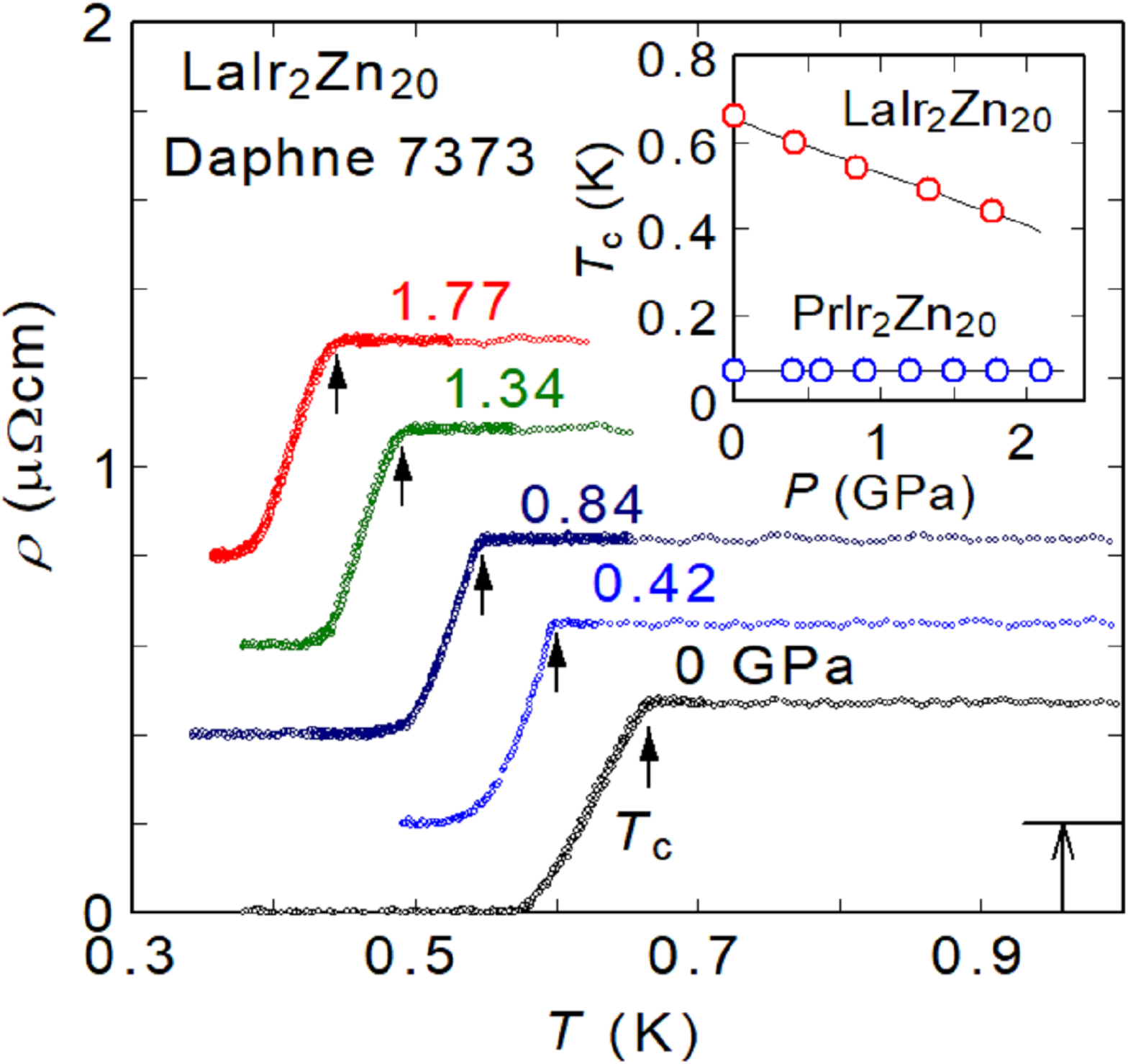}%
\caption{\label{fig:epsart} Temperature dependence of the electrical resistivity of 
LaIr$_{2}$Zn$_{20}$ under various constant pressures applied with Daphne oil 7373. Data sets are shifted upward consecutively by 0.2 $\mu \Omega $ cm for clarity. The inset shows the pressure dependence of the onset temperature $T_{\rm c}$ of the 
superconducting transition for LaIr$_{2}$Zn$_{20}$ and 
PrIr$_{2}$Zn$_{20}$.}
\end{figure}

\newpage

\begin{figure}
\includegraphics[width=10cm]{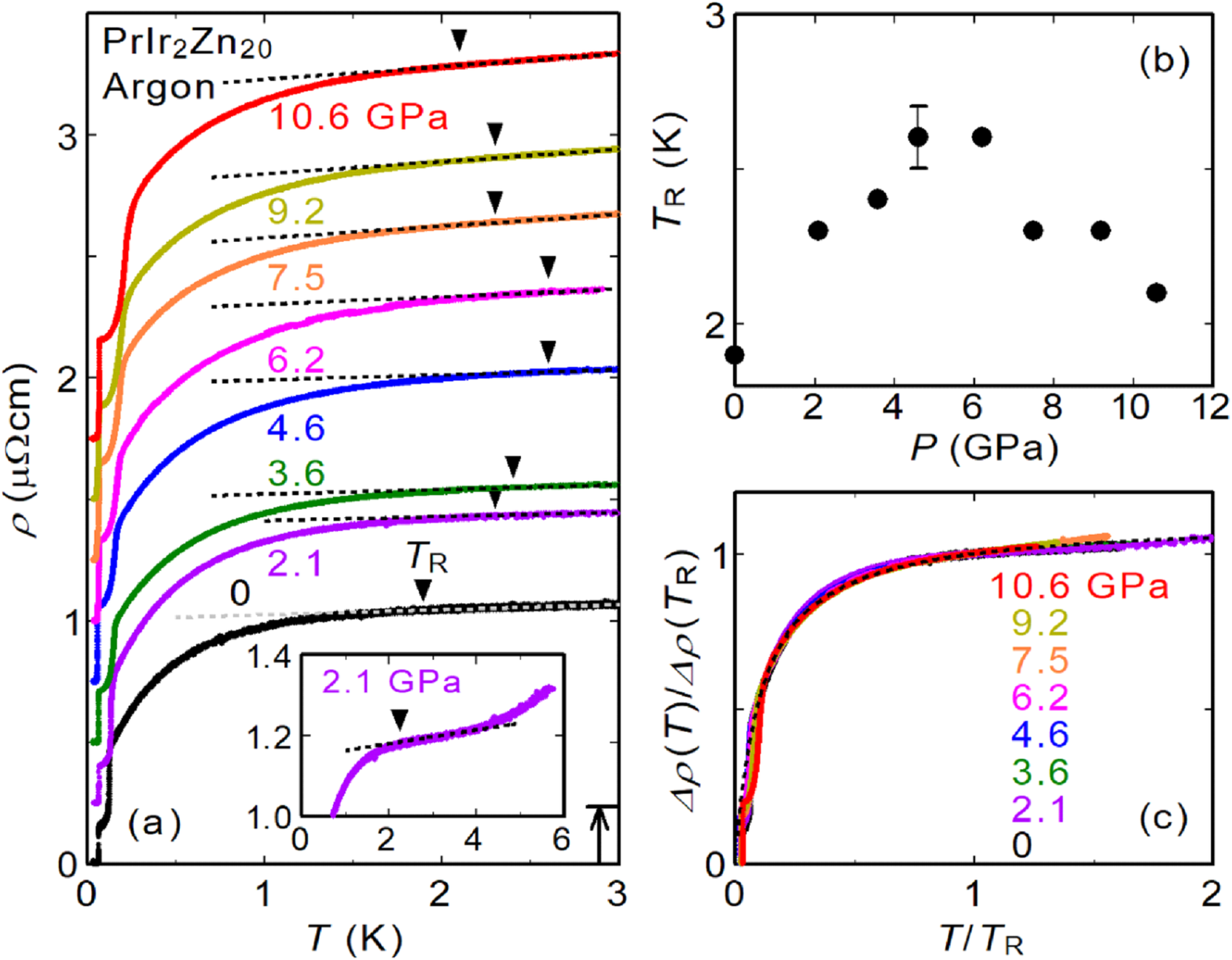}%
\caption{\label{fig:epsart}Temperature dependence of the electrical resistivity 
$\rho (T)$ of PrIr$_{2}$Zn$_{20}$ under pressures up to 10.6 GPa applied with 
argon as the transmitting medium. Data sets at various pressures are shifted 
upward consecutively by 0.25 $\mu \Omega$ cm for clarity. The arrows 
indicate the characteristic temperature $T_{\rm R}$, which is defined as the 
temperature where $\rho (T)$ starts deviating from the linear dependence on cooling. The inset shows the data of $\rho (T)$ at 2.1 GPa in the expanded temperature range.  (b) Pressure dependence of $T_{\rm R}$. (c) Scaling plot 
of the differential electrical resistivity $\Delta \rho = \rho (T) - \rho_{0}$ under various constant pressures. In the temperature region 
$0.1$ \textit{\textless T/T}$_{\rm R}$ \textit{\textless }1.2, $\Delta \rho$ follows the dashed curve calculated by using 
the two channel Anderson lattice model \cite{AT17}, as shown by the dashed curve. 
}
\end{figure}

\end{document}